\documentclass{emulateapj}
\usepackage{color}
\usepackage[normalem]{ulem}  
\usepackage{hyperref}

\newcommand{\dens}{g~cm$^{-3}$}

\slugcomment{To appear in ApJ ....}

\shorttitle{SN Ejecta-Companion Interaction Experiment Design}

\shortauthors{D. Garc\'\i a-Senz, P. Velarde, F. Suzuki-Vidal, C. Stehl\'e, M. Cotelo, D. Portillo, T. Plewa, A. Pak}

\begin{document}


\title{Interaction of hemispherical blast waves with inhomogeneous spheres: Probing the collision of a supernova ejecta with a nearby companion star in the laboratory}


\author{Domingo Garc\'\i a-Senz\altaffilmark{1,2}, 
Pedro Velarde\altaffilmark{3,4}, Francisco Suzuki-Vidal\altaffilmark{5},
Chantal Stehl\'e\altaffilmark{6}, Manuel Cotelo\altaffilmark{3,4},\\ David Portillo\altaffilmark{7,8}, Tomasz Plewa\altaffilmark{9}, Arthur Pak\altaffilmark{10}
}

\altaffiltext{1}{Departament de F\'\i sica, Universitat Polit\`ecnica de Catalunya, 
(EEBE) Av. Eduard Maristany 16, 08019 Barcelona, Spain; domingo.garcia@upc.edu} 
\altaffiltext{2}{Institut d'Estudis Espacials de Catalunya, Gran Capit\`a 2-4, 
 08034 Barcelona, Spain}
\altaffiltext{3}{Departamento de Ingenier\'\i a Energ\'etica, Universidad Politécnica de Madrid, 
 Jos\'e Gutierrez Abascal 2, 28006 Madrid, Spain; pedro.velarde@upm.es}
\altaffiltext{4}{Instituto de Fusi\'on Nuclear, Madrid, Spain}
\altaffiltext{5}{Imperial College London, UK}
\altaffiltext{6}{LERMA, Sorbonne University, Observatoire de Paris, PSL, CNRS, Paris, France}
\altaffiltext{7}{IMDEA Materiales, Madrid, Spain}
\altaffiltext{8}{ETSII, UPM, Madrid, Spain}
\altaffiltext{9}{Department of Scientific Computing, Florida State University, Tallahassee, FL 32306, USA}
\altaffiltext{10}{Lawrence Livermore National Laboratory, USA}



\begin{abstract}

Past high-energy density laboratory experiments provided insights into the physics of supernovae, supernova remnants, and the destruction of interstellar clouds. In a typical experimental setting, a  laser-driven planar blast wave interacts with a compositionally-homogeneous spherical or cylindrical target. In this work we propose a new laboratory platform that accounts for curvature of the impacting shock and density stratification of the target. Both characteristics reflect the conditions expected to exist shortly after a supernova explosion in a close binary system.

We provide details of a proposed experimental design (laser drive, target configuration, diagnostic system), optimized to capture the key properties of recent ejecta-companion interaction models. Good qualitative agreement found between our experimental models and their astrophysical counterparts highlights strong potential of the proposed design to probe details of the ejecta-companion interaction for broad classes of objects by means of high energy density laboratory experiments.

\end{abstract}




\section{Introduction}

Several important astrophysical problems involve collisions of strong shocks with diffuse or dense, quasi-spherical objects. This class of problems includes, for example, the interaction of a high-velocity supernova ejecta with the circumstellar medium, giving rise to supernova remnants (SNRs), the collision of shock waves with molecular clouds, and the impact of the blast-wave born in a supernova explosion with a nearby companion star. These events can be studied by means of astronomical observations, computer simulations and, more recently, scaled laboratory experiments. The synergy between these three seemingly independent methods is known as \textit{laboratory astrophysics}, and is a relatively recent approach aiming at shedding light on physical processes which are otherwise difficult to observe or simulate. For instance, the main features of the SNR dynamics have been successfully reproduced in the high-energy density laboratory experiments \citep{dra98,dra00} and showed good quantitative agreement with observations once appropriate scaling relations were applied \citep{ryu99}. In the context more closely related to the present work, the process of destruction of a spherical interstellar cloud was quite extensively studied in the laboratory \citep{kan01,han07,rob02}. These studies highlighted the central role played by hydrodynamic instabilities, mainly the Kelvin-Helmholtz (KH) and Widnall instabilities, in the cloud destruction.  

However, obtaining good agreement between laboratory experiments and more complex astrophysical objects may require more sophisticated initial configurations. In particular, problems involving self-gravitating objects may require considering density gradients. In the laboratory setting such gradients can be manufactured with help of layered spherical targets. Also, proximity between the shock source and the target object may demand accounting for curvature effects. For example, the supernova shock and the ejecta displays substantial divergence during the shock-envelope interaction in core-collapse supernovae. To study the impact of such effects on process of mixing in core-collapse supernovae, \cite{dra02} and \cite{gro13} proposed using diverging laser-driven shocks and multi-layered, hemispherical targets.

In this work we consider a laboratory experiment of the ejecta-companion interaction occurring shortly after a Type Ia supernova (SN Ia) explosion in the single degenerate (SD) scenario \citep{mar00,mao14}. In the proposed experiment, a laser-driven diverging, hemispherical blast wave collides with a layered, dense spherical target. In general, the present study offers the basis for future experiments relevant to physical scenarios involving self-gravitating objects that display significant density gradients. In particular, spherical two-layered targets ought to capture the basic characteristics of non-degenerate stars with their relatively rarefied envelopes and dense cores. 

The design details presented in this paper open up the possibility for future experimental realizations, as it has been the case of other proposals with different laboratory-astrophysics applications. For instance a novel platform to study magnetized accretions discs \citep{boc13}, and the design of experiments to study photo-ionization fronts \citep{dra16}.

One difficulty of the experiments is to produce scaled hemispherical shock waves instead of planar ones. Thus Section \ref{spherical} is devoted to characterize the impact of planar and hemispherical shock-waves onto inhomogeneous spheres. In Section \ref{astro} we describe the main features of the SD scenario of SNe Ia and present some large-scale simulations of the collision of the supernova ejecta with a nearby companion star. The insight obtained in the previous sections is used in Section \ref{laboratory} to devise and simulate a laboratory experiment to recreate the SD scenario, including detailed information concerning materials, laser energies as well as the geometrical setting of the experiment. The main results of our work are summarized and discussed in Section \ref{conclusions}.

\section{Spherical versus planar shock fronts impacting onto layered spheres}
\label{spherical}

In this section we want to asses the importance of the adopted geometry of a blast wave hitting dense, layered spheres. In this regard, the interaction between the SNR forward shock with cloud inhomogeneities in the interstellar medium have been the subject of several laboratory astrophysics studies in the past \citep{rob02,kle03,han07}. These experiments assumed planar shock fronts colliding onto homogeneous spheres, which are typically an order of magnitude denser than the environment. The assumption of planar shock geometry is adequate in this case because the supernova explosion originates in a point-like region located very far from the cloud. However, if the center of the explosion is not too distant from the object, curvature effects have to be taken into account to have a realistic depiction of the event. In this regard, it is worth noting the interaction of prompt supernova and nova blast waves interacting with a nearby companion star in compact binary systems or the bubble-bubble and bubble-surface interactions in fluid cavitation problems \citep{mae17}.

\begin{deluxetable*}{ccccrcccccc}[ht]
\tablecaption{Main features of the layered spheres used in the Toy-model experiments.}
\tablehead{
\colhead{Model}&\colhead{$M_{in}$}&\colhead{$M_{in}^c$}&\colhead{$E_{k}$}&\colhead{$E_{k}^c$}&\colhead{$\bar\rho_{blast}$}&\colhead{$\bar v_{blast}$}&\colhead{$v_{cm}^c=\frac{M_{in}^c~\bar v_{blast}}{M_c}$}&\colhead{$v_{cm}^c$~(hydro)} \cr 
 \colhead{-}&\colhead{g}&\colhead{g}&\colhead{erg}&\colhead{erg}&
 \colhead{\dens}&\colhead{cm~s$^{-1}$}&\colhead{cm~s$^{-1}$}&\colhead{cm~s$^{-1}$}}
\startdata
A&$0.435$&$0.1090$&$0.064$&$0.0160$&$1.660$&$0.542$&$0.0138$&$0.0149$ \\
B&$0.442$&$0.1105$&$0.065$&$0.0163$&$0.250$&$0.542$&$0.0140$&$0.0124$\\
C&$0.445$&$0.1112$&$0.064$&$0.0160$&$0.140$&$0.536$&$0.0139$&$0.0128$
\enddata
\tablecomments{$M_{in}, E_{k}$~are the mass and kinetic energy of the blast within the solid angle subtended by the outer radius of the envelope sphere and the explosion center. $M_{in}^c, E_{k}^c$~are the mass and kinetic energy of the blast within the solid angle subtended by the core. Columns 6 and 7 show the average density and average blast velocity. Column 8 gives the velocity of the center of mass of the core from momentum conservation and column 9 gives the hydrodynamic asymptotic estimate of the core velocity.} 
\label{ToyTable}
\end{deluxetable*}

\begin{figure}
\includegraphics[angle=0,width=1.0\columnwidth]{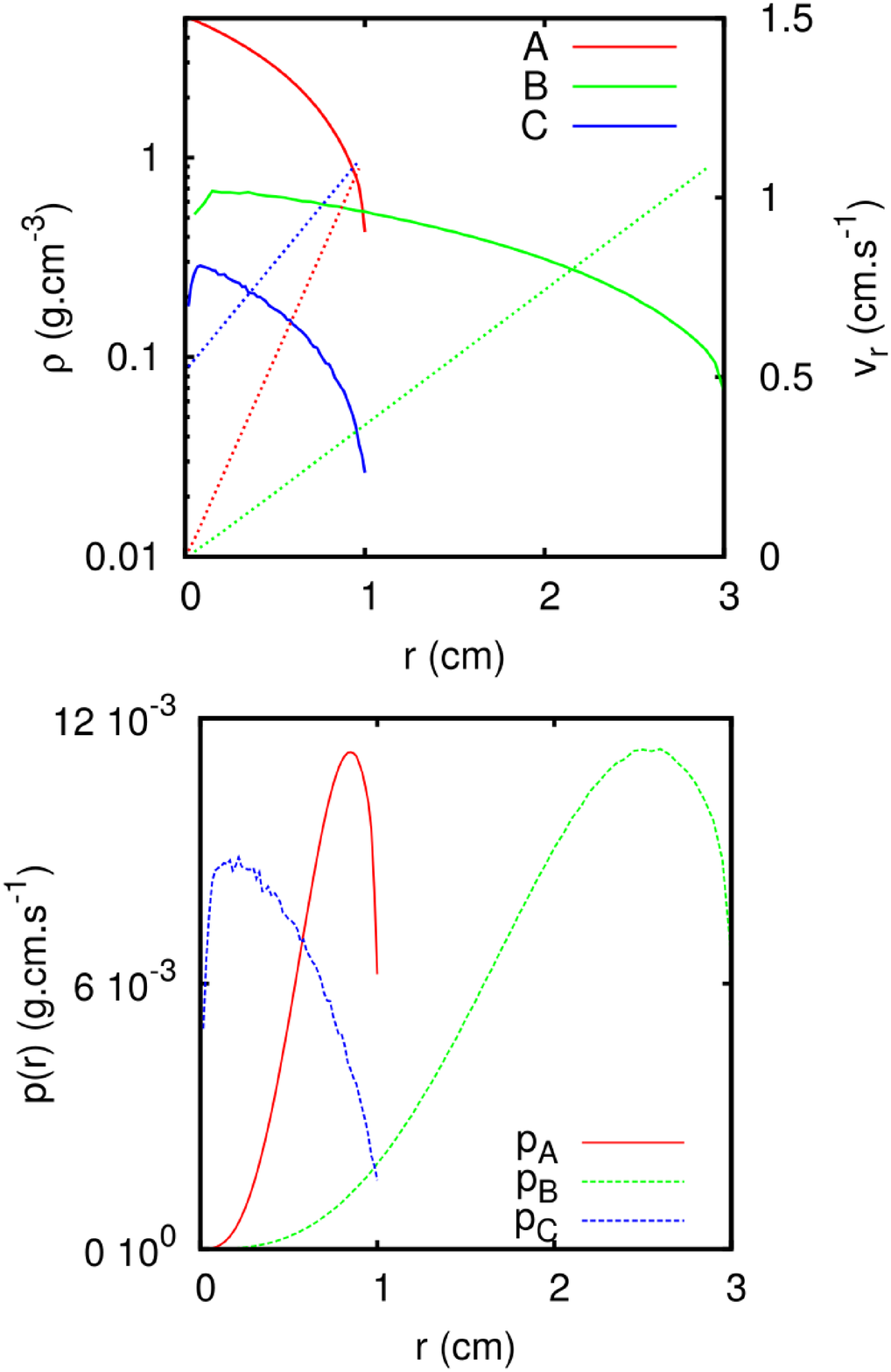}
\caption{Upper panel: Radial profiles of density (solid lines, log scale) and velocity (dotted lines, linear scale) of the blast-wave for models A, B, and C in Table \ref{ToyTable}. Coordinate $r=0$~is at the explosion origin (point-like in models A and B but planar in model C). Lower panel: Radial profile of the momentum of the blast-wave before the impact.}
\label{ToyAvRhoVel}
\end{figure}

To gain a general insight of the interaction of curved blast-waves colliding onto stellar laboratory analogues, a simple toy-model has been built. This model reduces the star to a two-layered sphere, the {\sl core} and the {\sl envelope}, characterized by their densities, $\rho_{c}, \rho_{e}$~and radius $R_c, R_e$. For illustrative purposes, the density and aspect ratio between core and envelope are taken as $\rho_c/\rho_e\simeq 50$, $R_e/R_c\simeq 2$~(equal core and envelope thickness) to mimic the structure of a Sun-like star (more details on the standard Sun model are presented in Section \ref{building}). The two-layer sphere is hit by blast waves with different curvatures: a) spherical 'close', with radius $R_{blast}=R_c+R_e$, (Model A in Table \ref{ToyTable}), b) spherical 'far', with $R_{blast}=3(R_c+R_e)$ (Model B) and c) planar (Model C). The impacting mass and kinetic energy had the same values in all three cases, and the equation of state of an ideal gas with $\gamma = 5/3$ was used. Table \ref{ToyTable} and Figure \ref{ToyAvRhoVel} summarize the values of the main parameters used in the simulations.

\begin{figure}
\includegraphics[width=1.05\columnwidth]{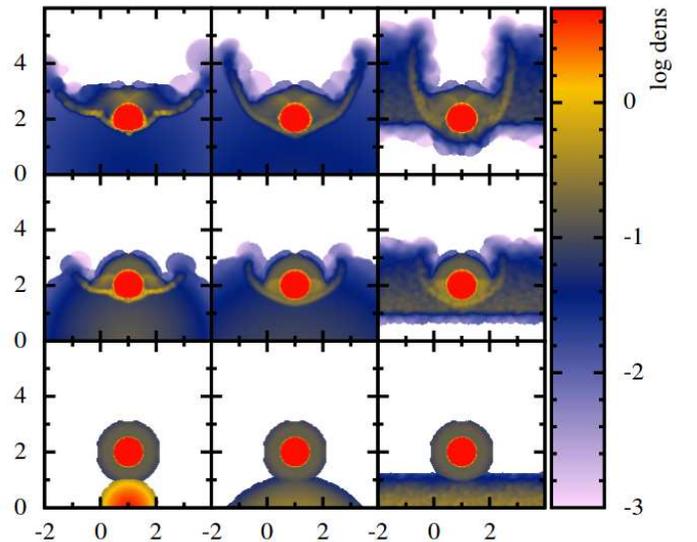}
\caption{Density color-map slice depicting the impact of blasts with different geometries (models A, B, C from left to right) onto layered spheres at three times, from bottom to top: t=0.0 s, t=2.0 s, and t=3.52 s (scale of each box in cm and $\rho$~in \dens).}
\label{FigSphynxDensToy}
\end{figure}

\begin{figure}
\includegraphics[width=1.05\columnwidth]{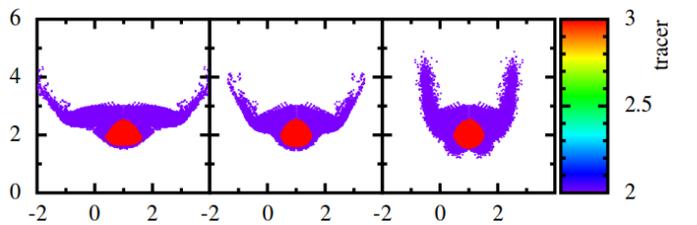}
\caption{Same as the top column in Figure \ref{FigSphynxDensToy} at t=3.52 s, but directly showing the distribution of the SPH particles belonging to the envelope material (characterized by the magnitude $tracer=2$, in blue) and the core ($tracer=3$, in red).}
\label{FigSphynxTracerToy}
\end{figure}

We have carried out several three-dimensional simulations of this toy-model with the cutting edge smoothed-particle hydrodynamics (SPH) hydrocode SPHYNX \citep{cab17}.  SPHYNX's most relevant feature is the use of an integral approach to handle gradients, which is more accurate than the standard SPH procedure. It also displays a better partition of unit, which reduces the tensile instability. Both improvements reduce the damping of short-wavelengths, thus leading to a better description of instabilities. Two-dimensional slices of the simulations are depicted in Figures \ref{FigSphynxDensToy} and \ref{FigSphynxTracerToy}. As it can be seen, curvature effects have a strong influence in the collision dynamics and thus cannot be disregarded. The planar front leads to a cylindrical-like shape in the post-shock ejecta-envelope region, whereas it becomes conical when the ejecta is spherical (Fig. \ref{FigSphynxDensToy}). Such effect is highlighted in Figure \ref{FigSphynxTracerToy} where a passive tracer was added to each component of the simulation to better track the  geometry of the stripped envelope. These results are reproduced by high-resolution simulations of the laboratory experiment using the code ARWEN, described in Section \ref{SimLab}. In the experiment, the choice of planar and spherical blast waves leads to very different angular distributions of stripped mass from a two-layered target sphere.

\begin{figure}
\includegraphics[angle=-90,width=1.0\columnwidth]{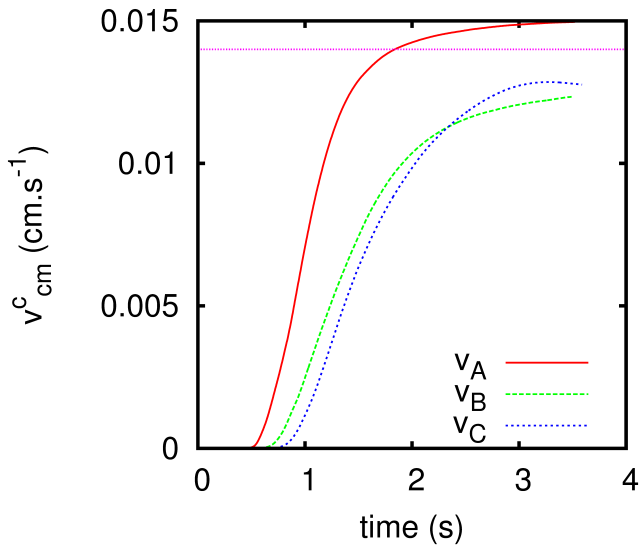}
\caption{Evolution of the velocity of the center of mass of the core for models A, B and C. The horizontal line is the core kick velocity obtained using momentum conservation for model A.}
\label{FigSphynxVelToy}
\end{figure}

The velocity of the center of mass of the core of the sphere from the toy model simulation for all three cases is shown in Figure \ref{FigSphynxVelToy}. The constant velocity represented by a horizontal line was obtained assuming that the core for case A is totally isolated and momentum is conserved. This agrees with the toy model result for case A (red line), with relative differences $\leq 13\%$. Also, the maximum relative ratio between models A, B and C is not large, $\le 20\%$, (see two last columns in Table (\ref{ToyTable}). These differences arise from the particular geometry of the blast-wave and the different density profiles in models A, B and C. The profiles of density and homologous radial velocity at t=0 s are depicted in Figure \ref{ToyAvRhoVel}, where the value $r=0$~cm is the explosion center. Although the functional  form of the adopted density profiles is the same (i.e.~linear with negative slope) the maximum and minimum values among them differ. In particular, the density values are larger in Model A, which lead to an enhanced piston effect. Thus, in spite of having a lower component of the velocity along the direction of impact, the core of model A gets the largest velocity kick. This is an indication that, besides the geometry, the particular density and velocity profiles of the blast-wave are also relevant to obtain the precise asymptotic velocity of the core. Figure \ref{ToyAvRhoVel} also shows the profiles momentum along the blast-wave for models A, B, and C. Despite the integrated total impacting mass, momentum and energy are the same in the three cases, their spatial profiles are not. 

Finally, it is worth noting that the evolution of models B and C is rather similar. This is not only due to the larger flatness of the blast-wave in model B, but also due to the lower differences in the maximum and minimum densities in their radial profiles with respect to model A (see Fig. \ref{ToyAvRhoVel}, upper panel)

\section{Connecting the impact of inhomogeneous spheres with the SD scenario of SNe Ia}
\label{astro}

 The quest for suitable progenitors of Type Ia supernova explosions is a long lasting hot topic of stellar astrophysics. Currently, two broad families of progenitors are considered. Both families involve a massive white dwarf (WD, the exploding object) but differ in the nature of the nearby companion star. In the so called Single Degenerate scenario (SD) the companion star can be a main sequence star or a more evolved star (a sub-giant or a giant star) whereas in the Double Degenerate scenario (DD) the secondary star is another white dwarf. In the SD scenario the companion star could survive the explosion of the WD (although severely perturbed) while in the DD scenario no remnant is left following the explosion. Therefore, the most straightforward way to distinguish between both scenarios would be to observationally detect the remnant of the secondary star. In this respect, although the first surveys of field stars around the central region of the Tycho SNR were promising \citep{rui04,rui14}, recent observations have cast doubts about the existence of such remnant of the companion star  \citep{xue15,wil16,rui18}. 

\begin{deluxetable*}{ccccrcccccc}
\tablecaption{Main features of the binary system at the explosion moment and at the beginning of the interaction, assumed as starting time t=0 s (last three columns). }
\tablehead{
\colhead{$M_{WD}$}&\colhead{$M_{Comp}$}&\colhead{Distance}&\colhead{$E_{kin}$}& \colhead{$\langle v_{eject}\rangle$}&\colhead{$t_{cross}$}&\colhead{$\Omega$} \cr
\colhead{$M_{\sun}$}& \colhead{$M_{\sun}$}&\colhead{$R_{\sun}$}&\colhead{$10^{51}$ergs}&\colhead{km.s$^{-1}$}&\colhead{s}&\colhead{str}}
\startdata
$1.36$&$1.0$&$2.95$&$1.28$&$6700$&$197$&$0.36$ \\
\enddata
\tablecomments{$t_{cross}$~is the time needed by the ejecta, moving at $\langle v_{eject}\rangle$, to cross the diameter of the secondary star with initial radius $0.95$~R$_{\sun}$. (reduced unit of time). $\Omega$~ is the solid angle subtended by the Sun-like star with respect to the center of the explosion.  } 
\label{table2}
\end{deluxetable*}


The identification of the companion star is the most direct, but not the unique way to confirm the SD scenario. During the interaction of the supernova ejecta with the companion, a big portion of the envelope of the companion star is stripped and the shocked supernova material mixes up with the envelope material of the secondary. The temperature in the interacting region rises to $T\simeq 6\times10^7$~K producing an excess of ultraviolet (UV) and soft X-ray transient emission, which has been detected \citep{cao15,mar16}. Also, the shielding effect of the companion star produces a low-density region in the, otherwise spherical, ejecta. Such kind of ejecta {\it hole}, is a geometric anomaly which may explain the current spectropolarimetric observations of SNe Ia \citep{kas04}. According to numerical simulations, the center of mass of the secondary receives a radial kick which raises its velocity from zero to $v_r\simeq 100$~km.s$^{-1}$~\citep{mar00,boe17}. In addition, the mixing between the supernova and the secondary material contaminates both objects, thus increasing the hydrogen content of the ejecta as well as the metal content of the surface of the remnant of the companion.

After some hundreds of years, the evolution of the ejecta hole could leave a fingerprint in the structure of the supernova remnant. Specifically, it could affect the geometry of the X-ray emission from the material swept by the reverse shock \citep{gse12, gra16}. The shadow cast by the companion star has been recently advoked as one of the possible scenarios to explain the asymmetric expansion of the Fe ejecta in the Kepler SNR \citep{kasu18}.        

The study of the impact of the ejected supernova shell with a nearby companion star is not new \citep{whe75}, although only after the work by \cite{mar00}, it has received a sustained momentum \citep{kas04,pak08,kas10,pan10,gse12}. Very recent works on the subject are those by \cite{gra16} and \cite{boe17}. The collision of a typical nova ejecta with a main sequence companion star was recently studied by \cite{fig18}, where the fate of the accretion disc surrounding the white dwarf was addressed.   

Lacking observations, the study of the interaction between the supernova debris and the companion star strongly relies in hydrodynamic simulations. For the most part, these calculations take advantage of the axisymmetric nature of the collision, and approach the phenomenon with 2D-cylindrical coordinates. Calculations in full three-dimensions are typically characterised by lower resolution but are able to better capture the growth of hydrodynamic instabilities (i.e.~Kelvin-Helmholtz, Ritchmyer-Meskhov or Widnall instabilities), as the seeds of these instabilities barely have a preferred geometry in nature. A 3D calculation could eventually allow the inclusion of orbital elements into the simulation.

\begin{figure*}
\includegraphics[width=\textwidth]{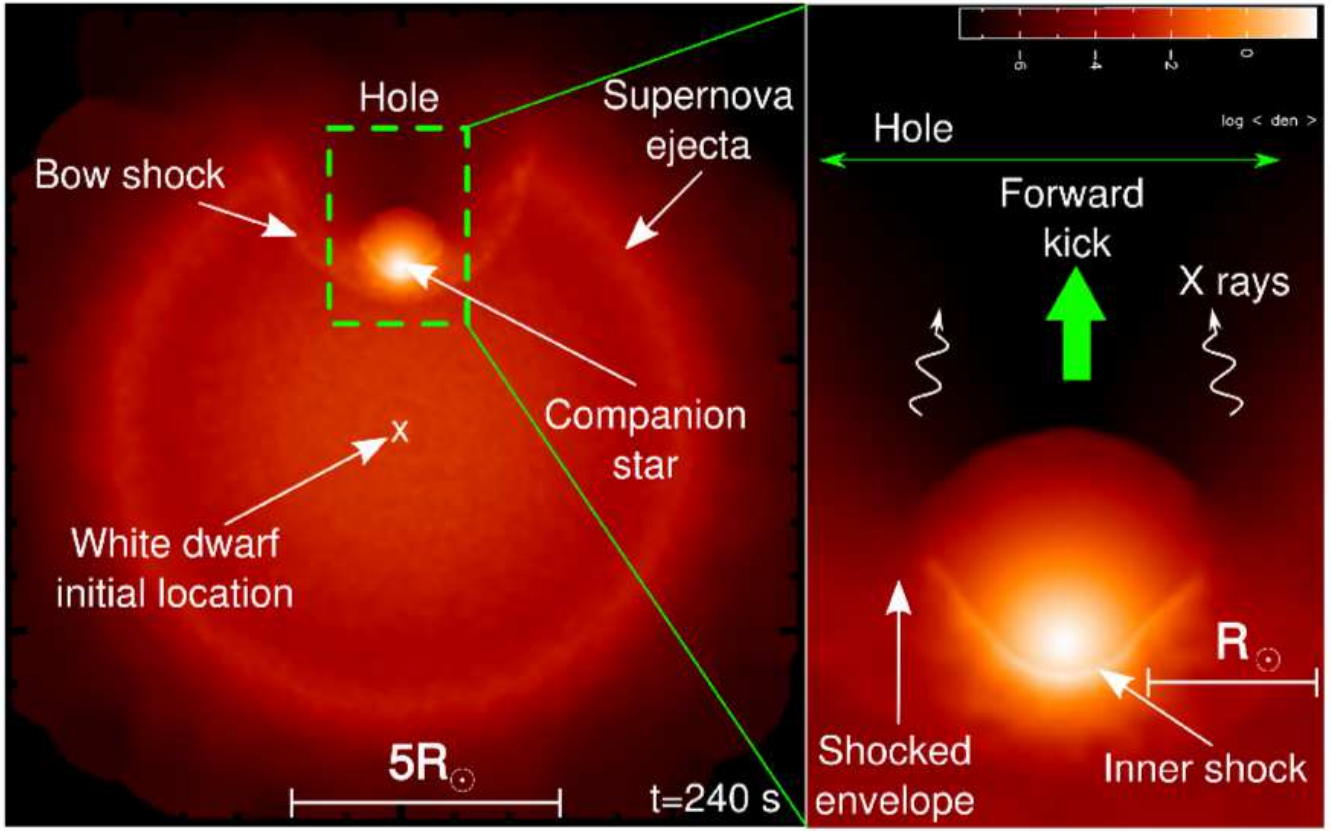}
\caption{Numerical simulations of the interaction of a supernova with a companion star with the 3-D hydrocode SPHYNX. The figures display 2-D slices in the collision plane at t= 240 s. Both images are density color maps showing  $\rho\ge 5\times 10^{-8}$~\dens~in logarithmic scale.} 
\label{FigAstro1}
\end{figure*}

%
%

A third route to study the collisional scenario described above is the laboratory-astrophysics approach. A controlled, reproducible laboratory experiment could shed light on many uncertain issues of the SD scenario and be complementary to hydrodynamic simulations. Ultimately, being able to probe the impact between the supernova ejecta and the companion star in the laboratory would aid in the search of the progenitor of the Type Ia supernova explosions. In the following section we describe the main features of the collision as obtained from numerical simulations of the astrophysical case from its start to around the first hour, once the interaction has ceased. 

\subsection{Hydrodynamics of the interaction}

\label{astrosimulations}

In the SD scenario the progenitor white dwarf is accreting mass from a nearby companion star. When the mass of the WD approaches the Chandrasekhar-mass limit, carbon begins to react with itself causing the explosion of the object \citep{nie00}. The companion star can be a normal solar-like star, a main sequence star with a larger mass, a sub-giant or even a red-giant star. An elemental widely used distance between the white dwarf and the companion star is set by the size of the Roche-lobe (RL) at the moment of the explosion. In the RL approach the geometrical features of the initial configuration of the system are, in first order, only function of the mass of both the WD and the companion star. We note, however, that the RL approach is a limit case and that, on average,  larger separations may be needed to avoid an excessive contamination of the supernova spectra with the hydrogen stripped from the companion star \citep{bot18}.  Table \ref{table2} summarizes the values of some relevant magnitudes at the moment of the explosion. For the companion star we have simply chosen the density, temperature and mean molecular weight of the standard Sun model \citep{bah04}\footnote{An extensive Table with detailed values of the standard Sun model used in \cite{bah04} are available at http://www.sns.ias.edu/~jnb/SNdata/solarmodels.html.}. The density, temperature, radial velocity and chemical composition of the Supernova ejecta was taken from a spherically symmetric hydrodynamical simulation of a Chandrasekhar-mass model of the explosion (model R2 in \cite{bra93}, with the homologous velocity rescaled so that the kinetic energy is $E_{kin}=1.28\times 10^{51}$~ergs). The explosion model was then mapped to a 3D distribution of $N_{eje}=2.0 \times 10^6$~SPH particles and put at a distance of $2.95$~R$_\sun$~from the center of the Sun-like star, which was in turn described with $N_{star}=1.5 \times 10^6$~particles.

\begin{figure*}
\includegraphics[width=1.0\textwidth]{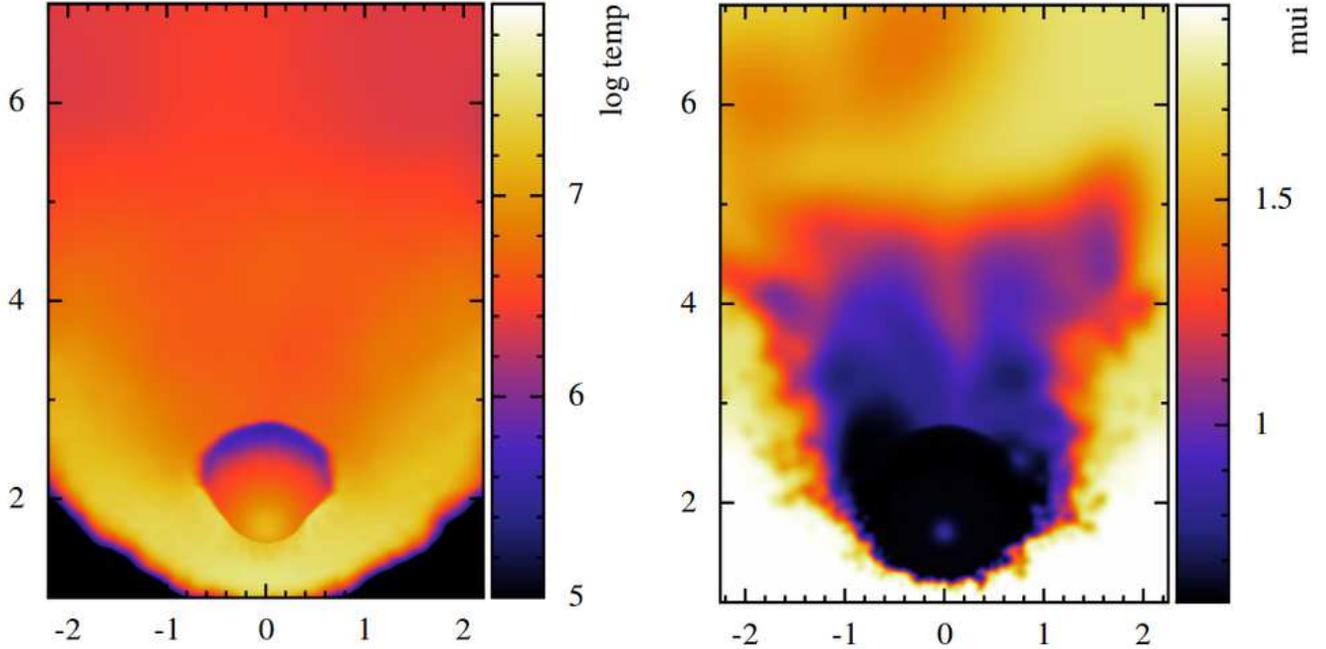}
\caption{Temperature (left panel) in Kelvin degrees and mean molecular weight ($\mu_i$, right panel) color maps of the interaction region at t=226 s. The stripped material from the companion has $\mu_i < 1$. The corrugation of the contact discontinuity (blue-red contact) between the ejecta high-$\mu_i$~material (yellow) and the low-$\mu_i$~material of the Sun-like star (blue) is due to the effect of the KH instability. Coordinates are in solar radius units.}
\label{FigAstro0}
\end{figure*}

The evolution of the system was studied in 3-D with the hydrocode SPHYNX. A representative snapshot of the interaction is depicted in Figure \ref{FigAstro1}. We can distinguish two shocks: namely the bow-shock embracing the companion star (Fig.\ref{FigAstro1}~left) and the innermost shock wave traveling through the interior of the Solar-like star (Fig.(\ref{FigAstro1})~right). The hole created by the companion, with an aperture $\simeq 40^{\circ}$, is clearly visible. The hypersonic collision heats-up both the ejecta and the envelope of the companion to $T\geq 10^7$~K (left panel in Figure \ref{FigAstro0}) which is observed as a transient - soft X-ray, UV - display lasting between minutes to hours \citep{kas10, bot18}.   

A scaled unit of time is obtained by dividing the diameter of the star, $D=2R_{\sun}$, and the average velocity of the homologous ejecta $ \langle v_{eject} \rangle$. Nevertheless, there is not an unique way to get $\langle v_{eject} \rangle$. It can also be estimated from the total mass and kinetic energy of the ejecta, which results in $\langle v_{eject}\rangle\simeq 9500$~km/s. But rather than an energy problem we are facing a momentum problem, so a more compatible estimation with the laboratory experiment is:

\begin{equation}
\langle v_{eject}\rangle \simeq \frac{\langle \rho(r)~v(r) \rangle}{\langle \rho(r)\rangle}
\label{avervel}
\end{equation}

\noindent
where $\rho(r)$~and $v(r)$~are the radial profiles of density and velocity in the spherically symmetric explosion model. This gives $\langle v_{eject}\rangle = 6700$~ km/s and a crossing time $t_{cross}= D/ \langle v_{eject}\rangle = 197$~s (Table 2). The maximum temperature is achieved by the shocked ejecta at the beginning of the collision ($T\simeq 10^8$~K) and decreases with time ($\simeq 5\times10^7$~K, $\simeq 3\times10^7$~K, $\simeq 10^7$~K, at $t=226$~s, $t=513$~s, and $t=1043$~s respectively). The shocked region radiates in the soft X-ray and UV bands.

The evolution of the contact discontinuity separating the ejecta from the star material is better addressed in Fig.(\ref{FigAstro0}) right, which depicts the distribution of the mean molecular weight $\mu_i$ at t=226 s. The mean molecular weight is a good tracer for this purpose due to its different value in the SN ejecta, $ \mu_i\geq 1.7$~and in the companion star, $0.62\leq\mu_i\leq 0.85$. As it can be seen, the contact discontinuity (blue-red layer in the lower half of the  figure) becomes corrugated and thus prone to develop shear-like instabilities, especially the Kelvin-Helmholtz (KH) instability \citep{boe17}. The growth of hydrodynamic instabilities around the shear layer might induce mixing between the metal rich supernova material with the  companion star, facilitating the contamination of the remnant of the companion with heavy elements. Nevertheless numerical simulations only show a low or moderate development of hydrodynamic instabilities around the shear region. Thus, the development of the KH and other instabilities does not seem to be very relevant to study the short term evolution of the collision. This is, however, a timely open question worth checking in a laboratory experiment. It goes in the same line as in precedent experiments on the Omega laser, which helped to disentangle the role played by the Widnall instability during the interaction of a shock wave with a solid sphere \citep{han07}.\\

\begin{figure}
\includegraphics[angle=-90,width=1.1\columnwidth]{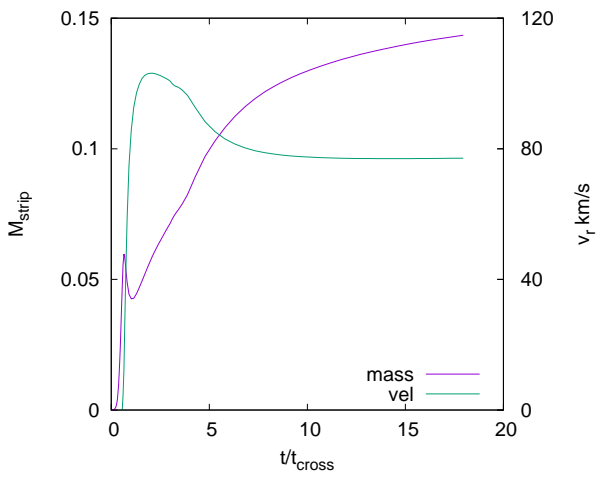}
\caption{Evolution of the stripped mass (in M$_\sun$) and radial velocity of the center of mass of the remnant of the companion star as a function of the normalized time. }
\label{FigAstro2}
\end{figure}

The evolution of the stripped mass and the asymptotic velocity of the remnant of the companion are shown in Fig.(\ref{FigAstro2}).  The values obtained with SPHYNX are robust as they are similar to those obtained by other groups \citep{mar00,pak08,boe17}, using very different hydrodynamic codes and semi-analytical estimates \citep{whe75}.       

\section{Proposed experimental platform}
\label{laboratory}


\subsection{Building the laboratory stellar analogue}
\label{building}

We use the results from the astrophysical simulation presented in the previous section to build up a simplified version of the companion star composed of two concentric layers with fixed densities. This will aid in the design of a laboratory stellar analogue that can mimic the overall properties of the companion star in the astrophysical case. In order to devise the model, we use the following definitions:\\

{\it Companion star (Astrophysics $\equiv A$)}: $\{M_{A2}$; $R_{A2}$; $M_{A1}$; $R_{A1}\}$. Where $M$ and $R$ stand for mass and radius, and the subscripts $2$~and $1$~refer to the companion before and after the collision respectively, i.e.~the subscript $1$~refers to the innermost region of the companion (the 'core') which survives to the collision.\\

\begin{figure*}
\includegraphics[width=\textwidth]{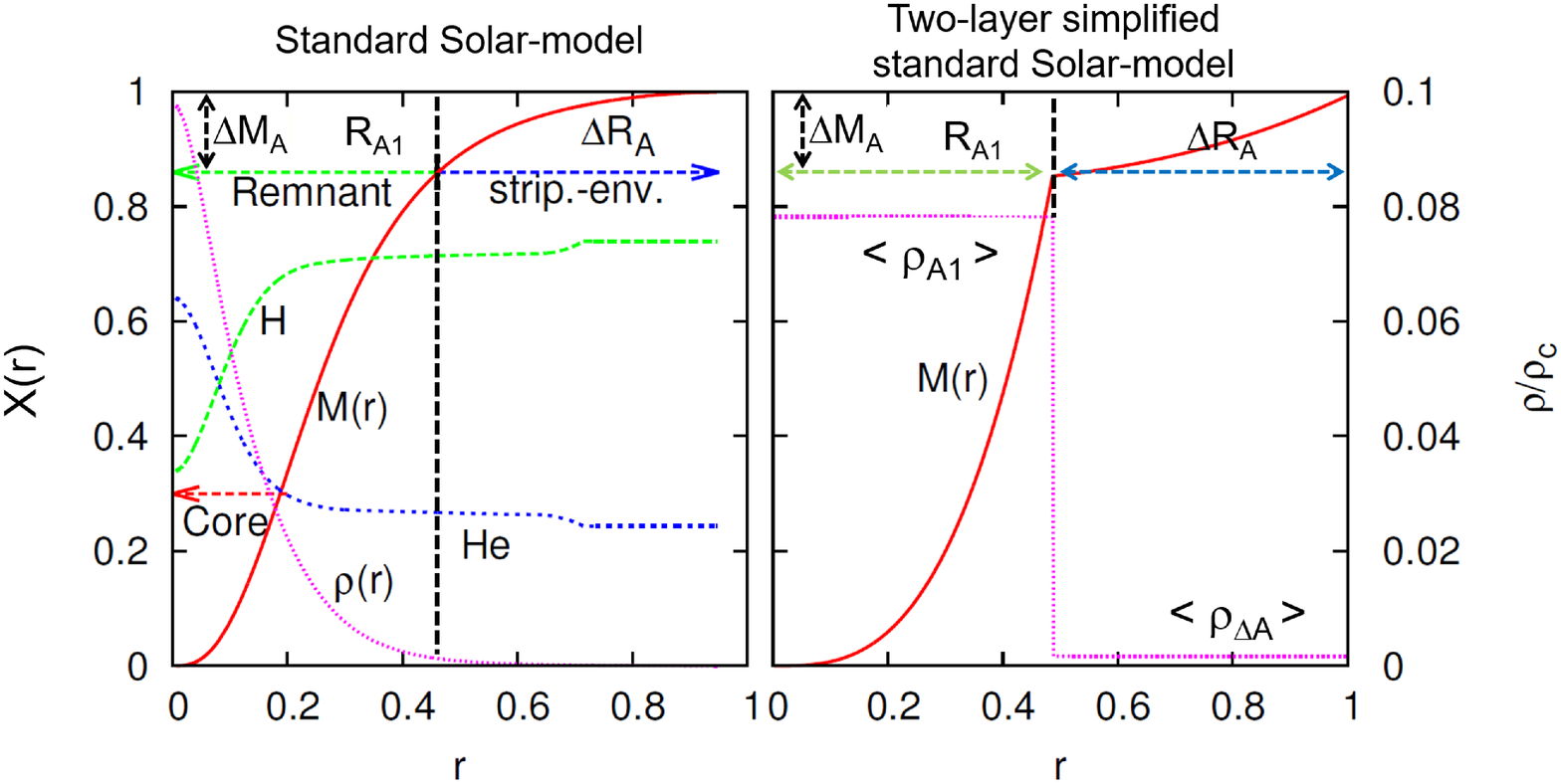}
\caption{Left panel: Profiles of several magnitudes, $\rho$(r), M(r), H(r), He(r) (generically labeled as X(r)) of the standard Sun model. The blue arrow indicates the width $\Delta R_A$ of the stripped-shell during the collision (inferred using the amount of stripped mass from detailed hydrodynamic simulations with the SPHYNX code), and the green arrow indicated the radius of the remnant companion $R_{A1}$. Right panel:  Equivalent simplified two-layer model of the companion star made of two averaged densities (normalized to the central density of the Sun, $\rho_c$ = 150 g~cm$^{-3}$) and equivalent radial mass distribution. This information is used to devise the two-layered laboratory target used in the experiment described in the text.}
\label{scaling1}
\end{figure*}

Radial profiles of mass and density for the companion from the standard Sun model before the collision are presented on the left panel in Figure \ref{scaling1}, with~$M_{A2} = 1 M_{\sun}$ and $R_{A2} = 0.95 R_{\sun}$\footnote{Note that in our numerical three-dimensional approach the radius of the companion is slightly smaller than the standard Solar radius}. The collision with the supernova ejecta results in mass loss from the outermost radius leading to the formation of a remnant core and an envelope with an amount of stripped mass $\Delta M_A= M_{A2}-M_{A1}$ and a thickness of $\Delta R_A=R_{A2}-R_{A1}$.

\begin{deluxetable*}{ccccrcccccccc}
\tablecaption{Scaling features of the companion star prior impact (except the last column which shows the asymptotic velocity of the remnant.}
\tablehead{
\colhead{$M_{A1}$}&\colhead{$M_{A2}$}&\colhead{$\Delta M_A$}&\colhead{$R_{A1}$}& \colhead{$R_{A2}$}&\colhead{$\Delta R_A$}&\colhead{$\langle\rho_{A1}\rangle$}&\colhead{$\langle\rho_{\Delta A}\rangle$}&\colhead{$f$}&\colhead{$g$}&\colhead{$h_A$}&\colhead{$AR_A$}&\colhead{$v_{A1}^{cm}$} \cr
 \colhead{$M_{\sun}$}& \colhead{$M_{\sun}$}& \colhead{$M_{\sun}$}&\colhead{$R_{\sun}$}&\colhead{$R_{\sun}$}&\colhead{$R_{\sun}$}&\colhead{\dens}&\colhead{\dens}&\colhead{-}& \colhead{-}&\colhead{-}&\colhead{-}&\colhead{km~s$^{-1}$}}
\startdata
$0.86$&$1.00$&$0.14$&$0.46$ &$0.95$ &$0.48$&$12.28$&$0.26$&$0.16$&$47.23$&$7.61$&$1.04$&$80.00$ \\
\enddata
\tablecomments{See Sections \ref{building} and \ref{scaling_relations}~for the meaning of the different magnitudes.}
\label{table3}
\end{deluxetable*}

The amount of stripped mass from numerical simulations found in the literature can take any value between $0.05 M_{\sun}\leq \Delta M_{A}\leq 0.3 M_{\sun}$, with the precise value depending on the mass and nature of the companion star, explosion energy and geometrical parameters \citep{mar00,boe17}. For the specific SPHYNX simulation presented in Figure \ref{FigAstro2}, the amount of stripped mass in the companion results in $\Delta M_{A} = 0.14 M_{\sun}$. We use this value on the mass profile from the standard Sun model to infer the radius of the core as $R_{A1}=0.46 R_{\sun}$ (i.e.~a thickness of the stripped shell of $\Delta R_A=R_{A2}-R_{A1}=0.49R_{\sun}$), with a core mass of $M_{A1} = M_{A2} - \Delta M_A = 0.86 M_{\sun}$. Notably a collision of this kind would shatter the whole convective zone of the Sun and a half of its radiative zone.



These values allow us to define fixed average mass densities for the core and the envelope for the simplified two-layer companion in the astrophysical case as:

\begin{equation}
\langle\rho_{A1}\rangle = \frac{M_{A1}}{\frac{4}{3}\pi R_{A1}^3}
\end{equation}
\begin{equation}
\langle\rho_{\Delta A}\rangle = \frac{\Delta M_A}{\frac{4}{3}\pi(R_{A2}^3-R_{A1}^3)}
\end{equation}

Radial profiles of average densities for the simplified two-layer companion are presented on the right panel in Figure \ref{scaling1}, with the main parameters shown in the first 8 columns of Table \ref{table3}.\\

\subsection{Scaling relations}
\label{scaling_relations}

We now use the simplified two-layer companion (right panel in Figure \ref{scaling1}) to device a scaled laboratory counterpart. As in the astrophysical case, we define the following parameters:\\

{\it Laboratory stellar analogue (Laboratory $\equiv L$)}: $\{M_{L1}$; $R_{L1}$; $\langle\rho_{L1}\rangle$;
$M_{L2}$; $R_{L2}$; $\langle\rho_{\Delta L}\rangle\}$. Similarly $\Delta M_L= M_{L2}-M_{L1}$~and $\Delta R_L=R_{L2}-R_{L1}$.\\

The physical and geometrical  features of the laboratory stellar analog can be obtained from the  following equalities

\begin{equation}
 f\equiv\left(\frac{\Delta M_A}{M_{A1}}\right)=\left(\frac{\Delta M_L}{M_{L1}}\right),  
 \label{mass}
\end{equation}

\begin{equation}
 g\equiv\left(\frac{\langle\rho_{A1}\rangle}{\langle\rho_{\Delta A}\rangle}\right)=\left(\frac{\langle\rho_{L1}\rangle}{\langle\rho_{\Delta L}\rangle}\right),
 \label{avdensity}
\end{equation}

Combining Eqs. (\ref{mass}) and (\ref{avdensity}) we get

\begin{equation}
R_{L2}=(1+h_A)^{\frac{1}{3}} R_{L1}
\label{radlab}
\end{equation}

with

\begin{equation}
h_A\equiv\frac{R_{A2}^3-R_{A1}^3}{R_{A1}^3}
\label{hscale}
\end{equation}


A characteristic unit of time is set by the crossing time, i.e.~the initial diameter of the companion star divided by the average velocity of the supernova ejecta, either in the astrophysical case or the laboratory,

\begin{equation}
t_{crossA,L} \equiv \frac{2 R_{A2,L2}}{\langle v_{A,L}\rangle}
\label{t_cross}
\end{equation}

\noindent
where $\langle v_A\rangle$=$\langle v_{eject}\rangle$=6700~kms$^{-1}$ for the astrophysical case (Eq.~(\ref{avervel})).\\ 



Similarly, the ratio of the terminal velocity of the center of mass (labeled with the superscript $cm$) of the sphere remnant after the impact and the mean velocity of the supernova ejecta can be scaled to the laboratory as,

\begin{equation}
\frac{v_{A1}^{cm}}{\langle v_{A}\rangle} =  \frac{v_{L1}^{cm}}{\langle v_{L}\rangle}
\label{velocity}
\end{equation}

\begin{deluxetable*}{ccccrccccccccc}
\tablecaption{Main features of the laboratory stellar analog prior impact after applying the scaling relationships in Table \ref{table3}~and assuming a solid $C_u$~sphere with a radius of $R_{L1}=252\mu m$.}
\tablehead{
\colhead{$R_{L1}$}& \colhead{$R_{L2}$}&\colhead{$\Delta R_L$}&\colhead{$\langle\rho_{L1}\rangle$}&\colhead{$\langle\rho_{\Delta L}\rangle$}&\colhead{$AR_L$}&\colhead{$D_L$}&\colhead{$v_{L1}^{cm}$} &\colhead{$v_{CH}^{cm}$} \cr
 \colhead{$\mu m$}&\colhead{$\mu m$}&\colhead{$\mu m$}&\colhead{\dens}&\colhead{\dens}&\colhead{-}&\colhead{$\mu m$}&\colhead{km~s$^{-1}$} &\colhead{km~s$^{-1}$}}
\startdata
$252$ &$515$&$263$&$8.9$&$0.19$&$1.04$&$1615$&$0.9$ & $55$\\
\enddata
\tablecomments{See Sect. \ref{scaling_procedure}, for the meaning of the different parameters. }
\label{table4}
\end{deluxetable*}


An additional geometrical constrain comes from the ratio of the initial radius of the companion $R_{A2}$ to the distance from the center of the supernova explosion $D_A$,


\begin{equation}
\frac{R_{A2}}{D_{A}} = \frac{R_{L2}}{D_{L}}
\label{curvature}
\end{equation}


The full scaling between the astrophysical case and the laboratory can be obtained from Equations (\ref{mass})-(\ref{curvature}). It is useful, however, to define the aspect ratio between envelope and core in both astrophysics and the laboratory

\begin{equation}
AR_{A,L} \equiv \frac{R_{A2,L2}-R_{A1,L1}}{R_{A1,L1}}
\label{aspectR}
\end{equation}


The main scaling parameters, estimated from the standard Sun model and the hydrodynamic simulations with SPHYNX, are presented in the last 5 columns of Table (\ref{table3}).\\

\subsection{Scaling procedure}
\label{scaling_procedure}

We propose to study the collision between a supernova ejecta and a nearby companion star in the laboratory by driving a supersonic, spherical blast-wave onto a solid, two-layer spherical target. A high-power laser is focused onto a concave, hemispherical cavity (a \textit{pusher}), which produces a curved blast wave. The proposed experiment is depicted schematically in the left panel of Figure \ref{target}.

To design the laboratory experiment, the density of the core of the two-layer spherical target $\rho_{L1}$ is chosen as a fixed parameter, while the radius of the spherical core $R_{L1}$ is left as a free parameter. Numerical simulations of the experiment are then performed with the 2-D adaptive mesh refinement (AMR) radiative hydrodynamics code ARWEN \citep{oga01} which allows refining the other parameters of the experiment.


The iterative procedure works as follows: with the trial value of $R_{L1}$~the total size of the target $R_{L2}$~is estimated with Eqs.~(\ref{radlab}) and (\ref{hscale}). Equation (\ref{curvature}) gives the distance between the pusher and the target center, $D_L$, while Equation (\ref{avdensity}) gives the density and mass of the envelope of the sphere $<\rho_{\Delta L}>$ and $\Delta M_L$ respectively. The terminal velocity of the core of the capsule $v_{L1}^ {mc}$ is then compared with the equivalent astrophysical velocity $v_{A1}^{mc}$ (Equation \ref{velocity}), and the trial value $R_{L1}$ is corrected adequately. The bisection process driven by $R_{L1}$~is repeated until a satisfactory consistency with the astrophysical and laboratory parameters is achieved.

As an example of a realistic material target choice, we take a core made of solid copper (density $\rho_{L1}$=8.9 g.cm$^{-3}$) with a radius $R_{L1}$=252 $\mu$m, the iterative procedure results in the values shown in Table \ref{table4}.\\



\subsection{Details of the experimental platform}
\label{details}

\begin{figure*}
\includegraphics[width=0.5\textwidth,angle=00,origin=c]{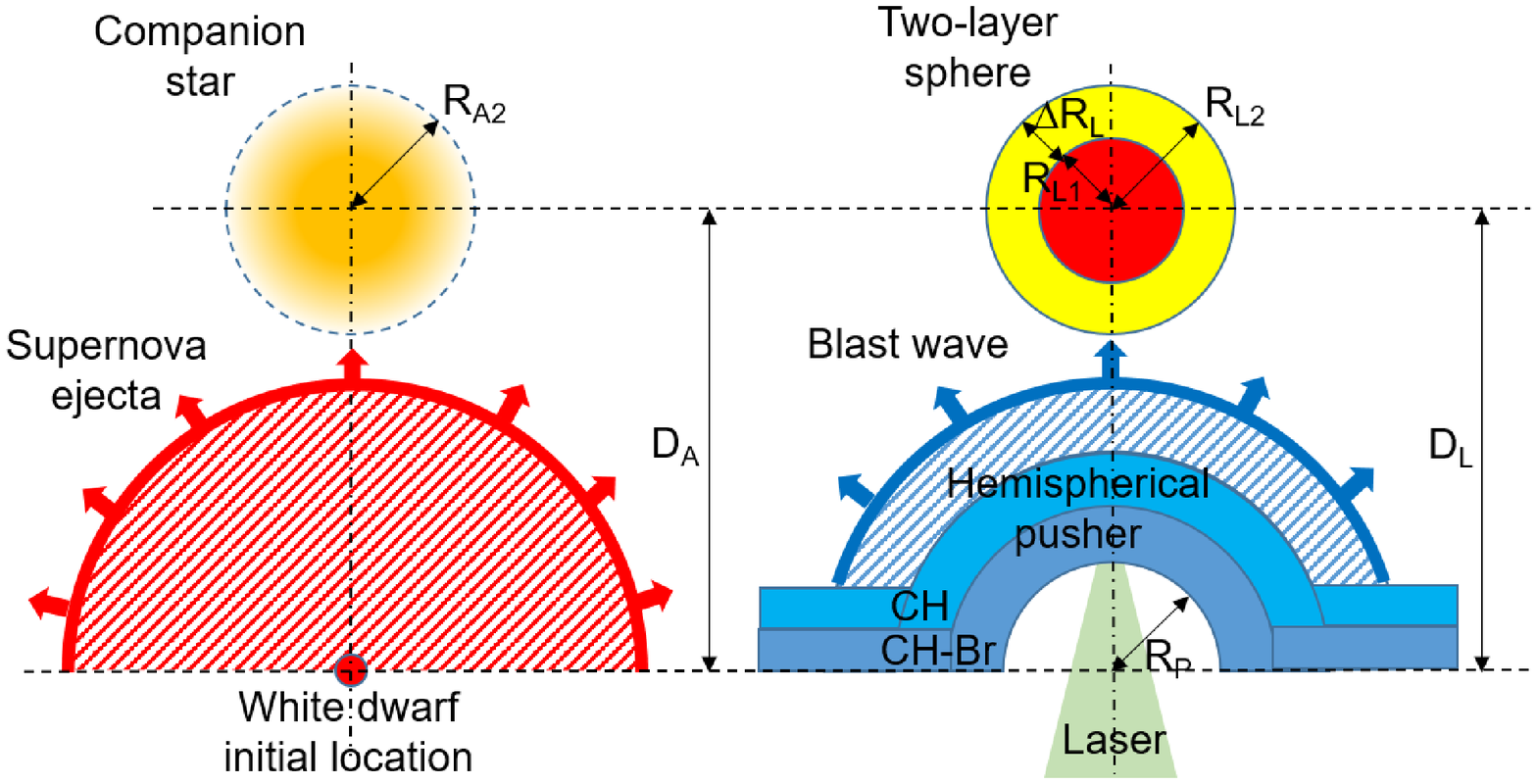}
\includegraphics[width=0.5\textwidth,angle=00,origin=c]{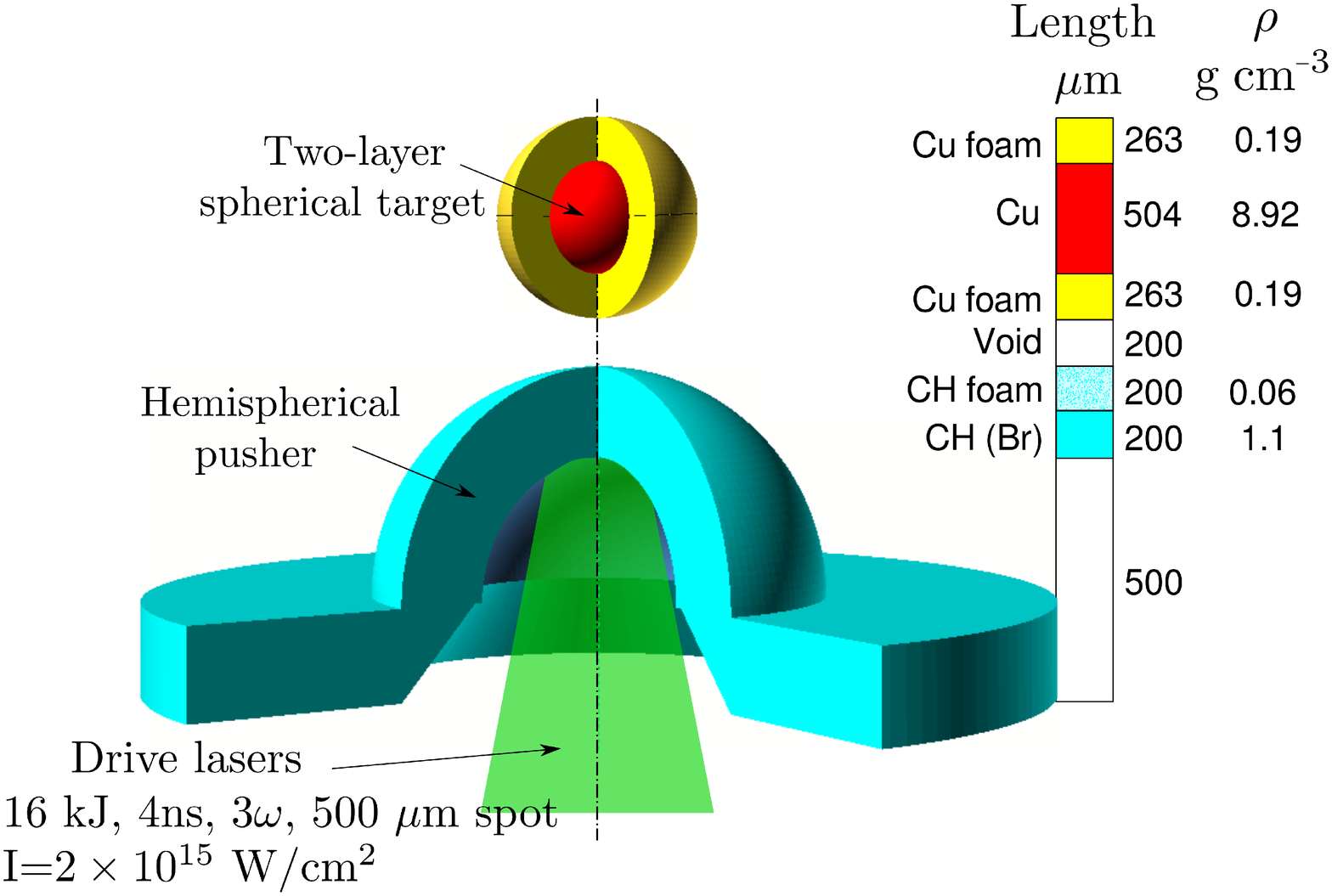}
\caption{Proposed experimental setup. Left: Schematic comparison between the astrophysical case and the laboratory platform. In the laboratory, a high-power laser is focused on the inside surface of a hemispherical pusher which drives a plasma flow onto a two-layered spherical target. Right: 3-D cut view of the experimental setup, showing the different materials, lengths and densities.}
\label{target}
\end{figure*}

The proposed experimental platform in Figure \ref{target} introduces two main differences compared to previous experiments (see e.g.~\citealt{kan01}). Firstly, the semi-hemispherical shape of the ablator introduces curvature to the blast wave in order to reproduce the interaction between the supernova and its nearby companion. Secondly, the two-layer structure of the spherical target allows to more realistically mimic the distribution of mass in the companion.

The hemispherical pusher has an inner radius of $R_P$=500 $\mu$m and is made of a layer of CH-Br (200 $\mu$m thick, $\rho=1.1$~\dens, in direct illumination with the laser) followed by a second layer of low-density CH foam (200 $\mu$m thick, $\rho=0.06$~\dens) which mimics the lower density envelope of the exploding white dwarf. The Br acts as a shield for the emission of X-rays from the interaction of the laser with the pusher. The two-layer spherical target has an overall diameter of $\sim$1 mm and is made of a solid copper core (diameter 2$R_{L1}$=504 $\mu$m, density $\rho_{L1}=8.9$~\dens) with a copper foam outer shell thickness $\Delta R_L$=263 $\mu$m and density $\rho_{\Delta L}=0.19$~\dens). The distance between the center of the target and the inside radius of the hemispherical pusher is $D_L$=1615 $\mu$m.

As in previous experiments with blast waves impacting onto spheres (see for example \citealt{kan01}), the blast wave emerging from the CH foam is separated from the target by a vacuum gap, which is set to $200~\mu $m in our case. Such empty region is wide enough so that the velocity profile of the blast in the pre-impact region becomes homologous (i.e.~velocity increases linearly with distance).

\begin{figure*}
\includegraphics[width=1\textwidth]{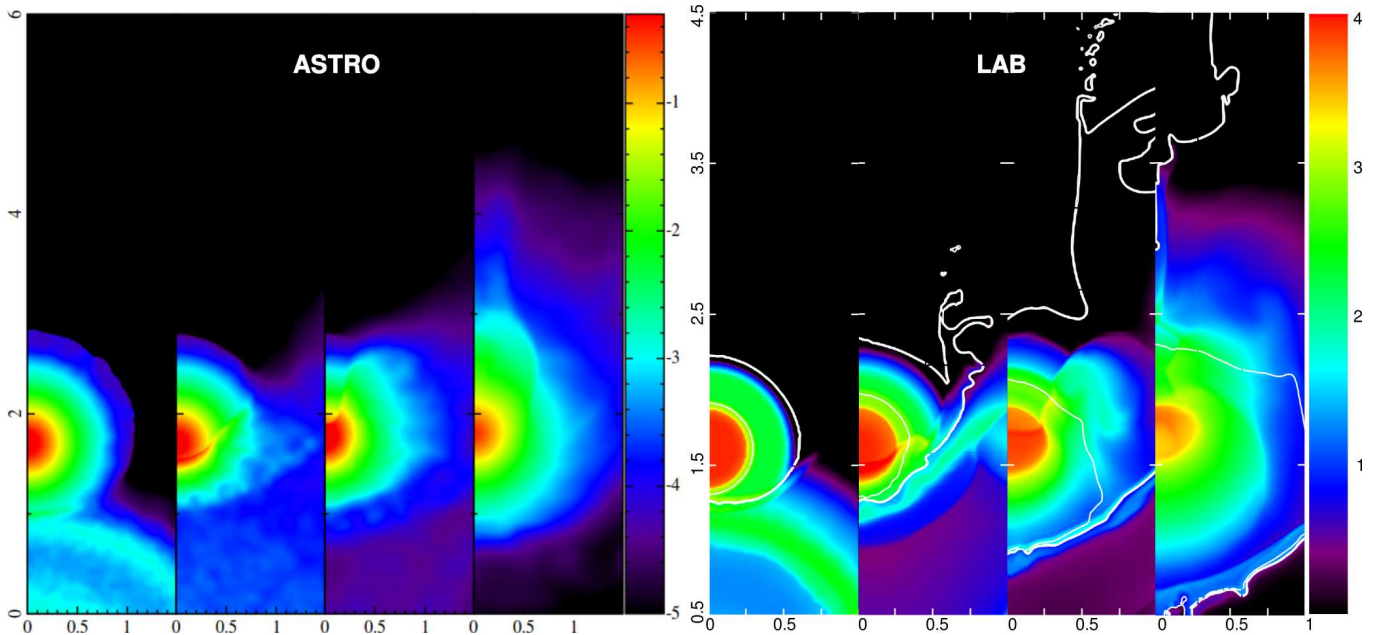}
\caption{Mass density (log scale) for the laboratory simulation with ARWEN (right, ``LAB') and a 2D slice of the 3D astrophysical simulation with SPHYNX (left, ``ASTRO'). The frames are at times $t/t_{crossA,L}$ = 0.3, 1.2, 2.6, 5.3, with $t_{crossL}$=19.4 ns for LAB and $t_{crossA} = 197$ s for ASTRO. White lines in the laboratory simulations with ARWEN represent the boundary between materials: Cu (solid), Cu (foam) and CH. Units are solar radius in ASTRO and mm in LAB.}
\label{Astrolab1}
\end{figure*}

As Figure \ref{target} suggests, the process by which the curved blast wave is produced is rather different in the astrophysical and the laboratory scenarios. In the supernova case the blast emerges from a point-like explosion whereas in the laboratory environment it comes after the laser ablation of a hemispherical cavity of finite radius. Nevertheless,  despite  these different generation mechanisms, the curvature of the blast is nearly the same in  both cases, as it can be checked by comparing  the ratio between the  radius of the target/star  and that of the blast-lab./blast-astro., (close to 0.5 in both cases) at the moment of the impact. The similarity in blast curvature is also seen in the comparison between the astrophysical and laboratory simulations  on the first panel in Figure \ref{Astrolab1}.  Additionally, a quantitative estimation shows that the surface radius of the laser-generated blast deviates from the spherical symmetry in less than $1\%$, when it hits the target.

\begin{figure}
\includegraphics[width=0.45\textwidth]{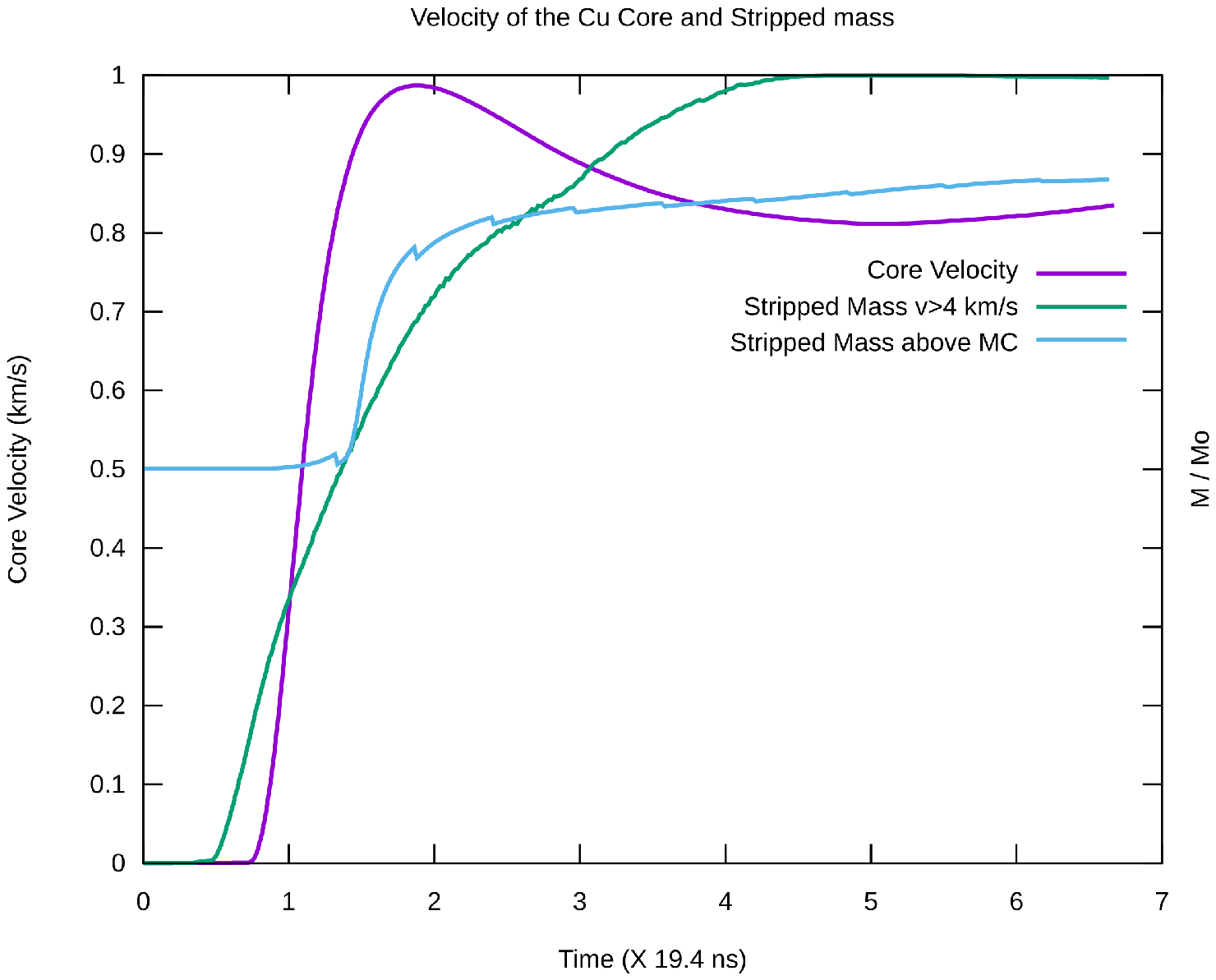}
\caption{Mean velocity of the spherical target copper core and stripped mass from the envelope as a function of time. Time is scaled with the mean pusher transit time ($\approx 19.4$ ns). The stripped mass is normalized with the total envelope mass and is identified as forward ($v_z>0$) traveling mass above the mass center (MC, blue line). Also plotted is the stripped mass obtained when its velocity becomes larger than 4 km/s (green line).}
\label{Astrolab3}
\end{figure} 

\subsection{Numerical simulations of the experiment}
\label{SimLab}
 
Results from numerical simulations of the scaled experiment with ARWEN are presented in the right panel of Figure~\ref{Astrolab1}. To drive the bow-shock, a laser with a total energy of 16 kJ, a wavelength of 451 nm (3$\omega$), a pulse duration of 4 ns, and a focal spot with a 500 $\mu$m diameter was used, thus resulting in an intensity of $\sim 2 \times 10^{15}$~W~cm$^{-2}$. These laser parameters are compatible with present high-power laser facilities, for instance one quad of the NIF laser \footnote{Kalantar, D., and Fournier, K. Introduction to NIF, LLNL-PRES-673980 (2016)}. Resolution is close to 5 $\mu m$ with a simulation box of 7 mm long. The code tracks all the initial material interfaces with negligible diffusion (even with the same thermodynamic properties) allowing the study of doping high-Z elements for experimental observation purposes. 

ARWEN is a 2D radiative hydrodynamics simulation code with adaptive mesh refinement (AMR) for both cartesian and cylindrical coordinates. This code couples an unsplit second-order Godunov method for non-diffusive and conservative multimaterial hydrodynamics, a flux limited diffusion package for electron heat conduction and a multi-group discrete-ordinate ($\rm S_n$) synthetically accelerated radiation transport module (\cite{garcia2010}). For the AMR structure, it uses the BoxLib\footnote{Current version AMRex \url{https://amrex-codes.github.io/}} package. ARWEN includes the physics required to handle laser ablation with intensities over  $10^{10} \ Wcm^{-2}$ and was designed to perform calculations of matter at High Energy Density conditions, which are common in laser-produced plasmas with laser intensities over $10^{14} \ Wcm^{-2}$. It has been applied to different types of systems involving laser-produced plasmas, from ICF (\cite{velarde_lpb2005}) to X-Ray secondary sources (\cite{oliva_hydrodynamic_2018}) or Laboratory Astrophysics (\cite{chaulagain_structure_2015}). We supply ARWEN with tabular Equation of State (EOS) and opacities to complete the model. The EOS used in the simulations presented here are based in QEOS (\cite{more_new_1988}) fitted to the available shock wave experimental data (\cite{cotelo_equation_2011}). For the opacities, we used the opacity code BigBART presented in \cite{de_la_varga_radiative_2011,de_la_varga_non-maxwellian_2013} for LTE conditions. We produce tables of the spectral mass absorption coefficient and latter collapse to the selected number of energy groups for the radiation transport package. BigBART uses the atomic physics code FAC (\cite{gu_fac_2008}) to compute self-consistent data such as oscillator strengths or radiative transition energies. The calculation with ARWEN presented in this article were performed with eight groups, six of the them with energies below 100 eV. We have run some control cases with sixteen groups to check the numerical results. In all the calculations we have used for the angular dependence of the intensity the $S_{6}$ approximation,i.e~6 directions for sampling an octant in the unit sphere at each spatial point (\cite{castor_2004}). In order to check the sensitivity of the results to the order of the angular approximation, some cases were run with $S_{10}$, i.e.~15 directions per octant.

Figure~\ref{Astrolab1} shows a time sequence of the collision from simulations of the laboratory and the astrophysical cases at the same scaled times $t/t_{crossA,L}$=0, 0.3, 1.2, 2.6 and 5.3. For the laboratory case this corresponds to times $t$=0, 5.8, 23.3, 50.4 and 102.8 ns after the laser pulse hits the pusher, whereas in the astrophysical case this corresponds to $t$=59, 236, 512 and 1044 s after the supernova explosion. Both simulations show very similar interaction dynamics, e.g.~an inner shock is driven through the spherical target/companion star (observed at $t/t_{crossA,L}$=1.2$-$2.6) which precedes the overall motion of the core of the companion as a 'kick'.

The overall dynamics of the interaction between the ejecta and the companion shows very little differences between both codes, suggesting that 3-D effects (i.e hydrodynamic instabilities, which eventually could show up in the astrophysical simulation) might not be important in the experiment.

Figure \ref{Astrolab3} shows profiles of the stripped mass from the target and the average core velocity as a function of the scaled laboratory time $t/t_{crossL}$ from simulations of the experiments with ARWEN. The evolution of the stripped mass and size of the Cu core remnant was estimated using two different methods. In the first method we take the criteria that any computational cell of the target is lost when its velocity exceeds a critical velocity, $v_{crit}^L = 4$~km s$^{-1}$. Such critical velocity is close to the escape velocity from the Sun surface, $v_{es}^A$~value, conveniently multiplied by the velocity scaling between astrophysics and laboratory (see Tables \ref{table2} and \ref{table4}), 

\begin{equation}
v_{crit}^L\simeq \frac{v_{CH}^{cm}}{\langle v_{eject}\rangle}~v_{es}^A
\label{velratio}
\end{equation}

The evolution of the stripped mass and core velocity obtained with this criteria is shown by the green line in Figure \ref{Astrolab3}. Our second procedure simply counts as stripped any envelope material with coordinates above the center of mass of the target. This second method leads to the blue line  in Figure \ref{Astrolab3}. As it can be seen, the precise criteria to decide when the mass element has been stripped mainly affects to the rising part of the curves, which is steeper with the second criteria. Nevertheless,  both criteria lead to a similar asymptotic behavior. The curves show a similar overall trend compared to the astrophysical case presented in Figure \ref{FigAstro2}. Furthermore, by making the ratio between the asymptotic core velocity  (AACV) and the average ejecta velocity (AEV) this results in a value of AACV/AEV$ \approx~1.2-1.6\times 10^{-2}$~for the astrophysical and the laboratory simulations respectively, which fulfills Equation (9). 

\begin{figure}
\includegraphics[width=1.05\columnwidth]{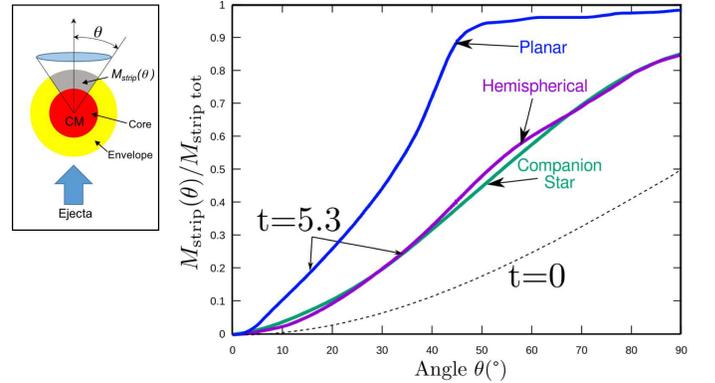}
\caption{Distribution of stripped mass of the companion star/two-layer target envelope as a function the azimuth measured from its center of mass (CM). The curve labeled 't=0' is the initial mass distribution before the collision. The curves labeled 'Companion Star' and 'Hemispherical' refer to the astrophysical and experiment simulations respectively. For comparison, the simulation of the experiment using a planar blast-wave generated with the same laser conditions (labeled as 'Planar') is also presented.}
\label{FigArwenComparisonPS}
\end{figure}

Figure \ref{FigArwenComparisonPS} shows a plot of the angular distribution of stripped mass from the 2-layer companion/target. The baseline curve labeled $t=0$~depicts the mass-distribution function of the Cu-envelope of the target before the impact (the shell in gray in the sketch) as a function of the axial angle $\theta$~ ($-90^0\leq\theta\leq 90^0$). Figure \ref{FigArwenComparisonPS} shows the laboratory and astrophysics  $M_{strip}~(\theta)$~ distributions at the common time $t/t_{crossA,L}=5.3$. The $M_{strip}(\theta)$~lines for the laboratory and astrophysical simulations lay above the reference line, indicating that much of the stripped material is coming from the lower hemisphere of the target. The consequences of using planar or spherical blasts in the laboratory setting are quite evident from the figure, and supports the conclusions with the simple toy model stated in Section \ref{spherical} above. For curved fronts the agreement between the astrophysical and laboratory simulations is excellent.


 




\section{Discussion and Conclusions}
\label{conclusions}

We presented the results of a proposed high-energy density laboratory experiment design 
in which a spherically expanding blast wave collides with a spherical target composited of a dense core surrounded by a low density shell. The experiment was optimized to match conditions expected to occur during the collision between a supernova ejecta with a non-degenerate, binary companion star in the Single Degenerate SN Ia formation scenario. The ejecta-companion collision is expected to produce observational signatures that could allow to constrain binary system parameters and the companion type, and thus advance our understanding of SNe Ia origins (cf.\ Section \ref{astro}). Several aspects of this problem could be investigated and verified in the laboratory.

The proposed experiment would enable validation of computer models through comparison of model predictions to experimental measurements in terms of properties of the low density region (hole) carved by the companion star in the SN ejecta, the companion velocity kick imparted by the ejecta, and the prompt X-ray emission produced by the shocked material. The experiment could also provide helpful information about the role of hydrodynamic instabilities in polluting the companion's envelope with the metal-rich SN ejecta material.      

The proposed design required addressing some challenging problems. The first issue was the construction of a laboratory analogue of the companion star. To this end, we represented a non-degenerate stellar companion as a spherical target composed of a dense central sphere surrounded by a shell made of lower density material. These two parts represented the companion's core and its envelope, respectively. As discussed in detail in Section \ref{building}, the physical properties of target components were carefully chosen so that their scaled values closely matched the average density and mass of the corresponding regions of the stellar model. Although the concept of a two-layered sphere may at first appear as a crude representation of the real star, it improves upon single-density targets typically used in laboratory experiments \citep{kan01,kle03}.

The second important design-related issue was the blast wave geometry. The use of planar wave fronts is justified when the distance between the explosion center and the target object is much larger than the target's radius. Unfortunately, this condition is not always satisfied in the case of SD SN Ia binary systems.
In the extreme case, the orbital distance is only few times the radius of the Roche-lobe filling companion. In this case, the diverging geometry of ejecta flow must be taken into account in order to correctly describe angular distribution of the stripped companion's envelope (cf.\ Section \ref{spherical}). Our simulation results indicate that curvature effects become negligible if the orbital distance is greater than about 5 times the companion radius.

Figure \ref{target} shows our experimental design configuration which accounts for both a composite density structure of the target and the blast wave divergence. The blast wave is driven by laser-ablation of a hemispherical cavity (a pusher) made of plastic, whereas the target consists of a copper sphere (the core) surrounded by the shell made of copper foam (the envelope). The proposed composite target is made of readily available materials and can be produced using currently available target manufacturing technology.

The simulation results of the experimental system compare very favorably to computer models of the ejecta-companion interaction. The angular distribution of the stripped target envelope material closely matches that of the envelope of the companion star. Also, the evolution of the velocity of the center of mass of the remnant is qualitatively similar in both cases. The asymptotic limit velocity of the remnants of the stellar analog and companion star agree, after scaling, within $30\%$. The impact of the blast wave is expected to impart a $\sim$0.9 km/s velocity \textit{kick} onto the core of the surrogate star.  This velocity could potentially be measured via the NIF VISAR diagnostic, which can diagnose velocities as low as 0.5 km/s. Future work will investigate what modifications to the target geometry would be required, such as a reentrant cone, to incorporate a VISAR measurement.


\begin{figure}
\includegraphics[width=1.05\columnwidth]{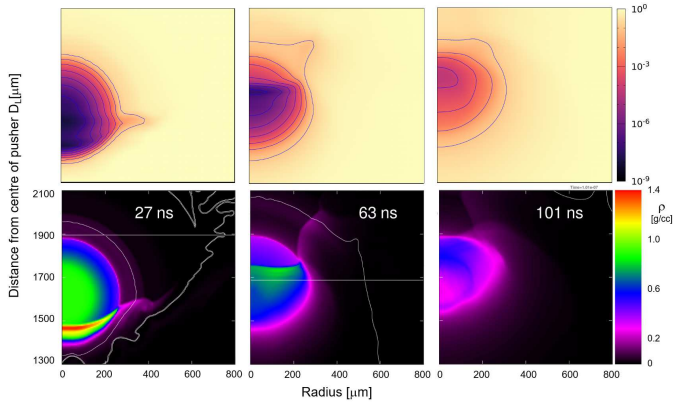}
\caption{Synthetic X-ray radiography from ARWEN simulations of the laboratory stellar analogue at scaled times $t/t_{\mathrm cross~L}= 1.39, 3.25$~and $5.21$~respectively.The bottom row shows mass density for each time. The top row shows the transmission coefficient for 18 keV photons (Mo backlighter), with isocontours corresponding to a factor of 10 change in transmission.}
\label{Radiographypng}
\end{figure}

\begin{figure}
\includegraphics[width=1.00\columnwidth]{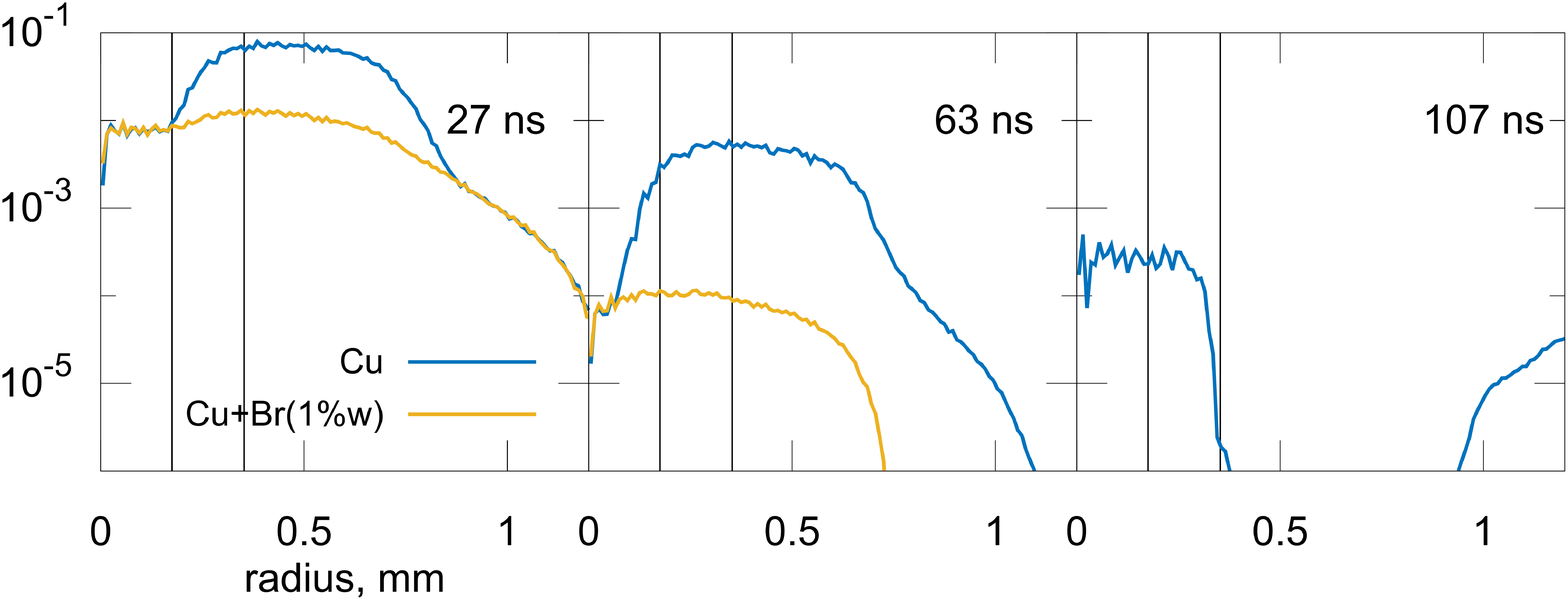}
\caption{Simulated plasma self-emission as a function of radius from the two-layer spherical target at the same times as Fig.~\ref{Radiographypng}. The signal (arbitrary units) was averaged in the spectral range 50~eV to 150~eV and was produced by a $50\ \mu m$ pinhole placed $4.385$ mm away from the center of the core of the target together with a screen 3 mm away from the pinhole along pusher-target axis. For each time, the emission from a pure Cu target and a Cu envelope doped with Br are compared. The emissivity at $107\ ns$ is negligible for the doped target. The two vertical lines indicate the original radius of the target core and envelope.}


\label{Emissivitypng}
\end{figure}


Our next goal is to experimentally realize the proposed setup, and begin to acquire experimental data to compare with our simulation results. The presented model experimental results were obtained using laser parameters currently achievable at large-scale laser facilities, such as the National Ignition Facility (NIF). The proposed experiment requires a modest laser drive energy of 16 kJ delivered by two NIF quads over 4 ns with a top-hat temporal profile. The drive beams are smoothed with phase plates and the spot has a super-Gaussian spatial profile with a 1000 $\mu$m FWHM. 

The target evolution is diagnosed using an X-ray backlighting imaging technique. The system uses 18~keV quasi-monochromatic X-rays emitted by a Mo backlighter energized with 36 kJ of delivered in 1 ns by four NIF quads. The diagnostic system was optimized with help of synthetic diagnostics in which hydro simulation results are post-processed using a ray-tracing method. The ray tracer calculates ray paths for a predefined number of rays with their initial intensities calculated using the local plasma emissivity. Ray trajectories are integrated with their direction changing according to the local value of the plasma refraction index while their intensities are attenuated due to plasma absorption.

The synthetic radiographic images are shown in Figure \ref{Radiographypng}. At early times (left pair of panels in Figure \ref{Radiographypng}), the low density target envelope is only partially overrun by the shock, which has not reach the dense target central core region yet.
As time progresses, the incoming flow completely engulfs the target. The shock wave loses strength  as it moves into the Cu core, but shocked Cu has high enough density to remain completely opaque to diagnostic radiation at all times.

Because prompt X-ray and EUV emission are one of the key predicted observational signatures of the ejecta-companion interaction \citep{kas10,bot18}, it is interesting to consider self-emission of the shocked plasma in the corresponding experimental system. Figure \ref{Emissivitypng} shows the radial distribution of the model target emission averaged in the spectral range between 50~eV to 150~eV, at the same elapsed times as in Figure \ref{Radiographypng}. The emission maps were obtained using a pinhole camera located 6~mm away from the center of the pusher (i.e.~a distance of 4.385~mm betweeen the centre of the spherical target and the pinhole), with the imaging plane placed at a distance of 3 mm from the pinhole. The first two panels of Figure \ref{Emissivitypng} show the radial distribution of the emission from the target envelope made of either pure copper or copper doped with $1\%$ bromine. The two target types appear qualitatively and quantitatively different. The pure copper target emission is composite in appearance with the emission from the target envelope dominating at early and intermediate times while the core is the only source of emission at late times. Initially, the emission in the brominated target is rather uniform, with the emission source decreasing in size as time goes on. This suggests that one can obtain more detailed information about shock evolution through the target by using various doping agents, possibly with two or more doped layers.  

Furthermore the self-emission and radiography data can also potentially provide 
additional information about the extent of hydrodynamic mixing between the envelope and core regions and with the pusher material. Imaging the self-emission from multiples lines of sight could allow for the reconstruction of the stripped mass distribution and cone angle opening.


We note that, compared to predictions of astrophysical models \citep{bot18}, the target self-emission significantly differs in terms of its spatial distribution and temporal behavior. One possible reason for this difference is the stronger density stratification of stellar envelopes, with the shock energy quickly thermalized in the outermost envelope layers. The prompt model emission is also much harder in the case of astrophysical system due to much higher transmitted shock speed and the corresponding temperatures of the shocked envelope material.

\acknowledgements

The authors would like to thank Marius Millot for useful discussions regarding the VISAR technique.
This work has been supported by the MINECO Spanish project AYA2017-86274-P and The Generalitat of Catalonia SGR-661/2017 (DG).
PV and MC acknowledge support from the VOXEL project, funded under European H2020-FET Open research projects (ID 665207).
FSV acknowledges support of The Royal Society through a University Research Fellowship.
AP acknowledges support from the U.S. Department of Energy (DOE) under Contract No. DE-AC52-07NA27344.
DG acknowledges the support and hospitality provided by the Plasma Physics Group at Imperial College London, during the writing process of this manuscript. 
This research used resources of the National Energy Research Scientific Computing Center, a DOE Office of Science User Facility supported by the Office Of Science of the U.S. Department of Energy under Contract No. DE-AC02-05CH11231.

\end{document}